\documentclass[submission,copyright,creativecommons]{eptcs}
\providecommand{\event}{PC2016} % Name of the event you are submitting to
\usepackage{breakurl}             % Not needed if you use pdflatex only.

\usepackage[pdftex]{graphicx}				% Use pdf, png, jpg, or eps with pdflatex; use eps in DVI mode
\usepackage{amssymb}
\usepackage{amsmath}
\usepackage{braket}
\usepackage{tikz}
\usepackage{pgfplots}
\usepackage{filecontents}
\usepackage{hyperref}
\usepackage[makeroom]{cancel}
\usepackage{turnstile}
\usepackage[hypcap]{caption}
\usetikzlibrary{arrows, decorations.markings, matrix}

\usepackage{bussproofs}
\EnableBpAbbreviations
\newcommand{\seq}{\Rightarrow}

\newcommand{\RLab}{\RightLabel}
\newcommand{\LLab}{\LeftLabel}
\usepackage{amsthm}
\newcommand{\logicname}{$\mathbf{QMC}$}
\newcommand{\qax}{\ket{0} \seq}
\newcommand{\hadz}{\frac{1}{\sqrt{2}}\ket{0} + \frac{1}{\sqrt{2}}\ket{1}}

\title{Sequent Calculus Representations for Quantum Circuits}
\author{Cameron Beebe
\institute{Research Center for Neurophilosophy and Ethics of Neurosciences, \\ Graduate School of Systemic Neurosciences, \\ LMU Munich, Germany}
\institute{Munich Center for Mathematical Philosophy, \\ LMU Munich, Germany}}
\def\titlerunning{Sequent Calculus Representations for Quantum Circuits}
\def\authorrunning{C. Beebe}
\begin{document}
\maketitle

\begin{abstract}
When considering a sequent-style proof system for quantum programs, there are certain elements of quantum mechanics that we may wish to capture, such as phase, dynamics of unitary transformations, and measurement probabilities.  Traditional quantum logics which focus primarily on the abstract orthomodular lattice theory and structures of Hilbert spaces have not satisfactorily captured some of these elements.  We can start from `scratch' in an attempt to conceptually characterize the types of proof rules which should be in a system that represents elements necessary for quantum algorithms.  This present work attempts to do this from the perspective of the quantum circuit model of quantum computation.  A sequent calculus based on single quantum circuits is suggested, and its ability to incorporate important conceptual and dynamic aspects of quantum computing is discussed.  In particular, preserving the representation of phase helps illustrate the role of interference as a resource in quantum computation.  Interference also provides an intuitive basis for a non-monotonic calculus.  
\end{abstract}

%\section{New Notes}

%\newpage
%\tableofcontents

%\newpage
\section{Introduction}

In Gentzen-style sequents for structural proof theory, for example the system \textbf{G1cp}, elements in the antecedent (represented by the multi-set $\Gamma$) are interpreted as having a conjuctive relationship to one another, whereas elements in the succedent (the multi-set $\Delta$) are interpreted as having a disjunctive relationship between one another.  In other words, on the `meta-level' we read the sequent $\Gamma \seq \Delta$ classically\footnote{For now, the sequent arrow $\seq$ is equivalent with $\vdash$, although later there will be a distinction between the two symbols.} as
\begin{equation}
\bigwedge (\Gamma) \to \bigvee (\Delta).
\end{equation}

When writing out some particular proof of an argument, the elements of $\Gamma$ and $\Delta$ are separated by commas (i.e. $\Sigma, A, A\to B \seq B, \Theta$).   These `structural' commas between the terms are read in the metalanguage as `$\land$' on the left side, and as `$\lor$' on the right side.  From the conjunction of the `premises' in the antecedent, we can obtain proofs of the `conclusions' in the succedent.  Then, a truth-functional interpretation of the sequent corresponds classically in that the sequent holds iff at least one element in $\Gamma$ is false or one of $\Delta$ is true (see e.g. Paoli \cite[\S 3.1]{Paoli2002}). 

In the sequent calculus \textbf{G1cp}, in addition to the operational rules for the various connectives $\{ \land, \lor, \to, \lnot \}$ we include the structural rules of (left and right) weakening and contraction, and can show the admissibility of $Cut$.   Various so-called ``sub-structural logics'' take aim at these structural rules, for example we could have reasons for dropping weakening, contraction, or $Cut$ from a logical system.  An intuitionistic system such as \textbf{G1ip}, for example, will restrict proofs to one term (or no term) in $\Delta$ in the succedent---denying the general ability to introduce formula by right weakening\footnote{Unless, of course, $\Delta$ is empty.} and therefore also we will never be able to eliminate a duplicate occurrence of a formula by right contraction.

Other examples are found in the literature concerning a quantum logic, or a logic that is based on theoretical or mathematical considerations from quantum physics.  Unfortunately, there are many problems to be found in the traditional motivations and attempt at defining a satisfactory quantum logic---not to mention problems in explicating why exactly these logical systems are important, or how they are related to the physical world.\footnote{For example, consider that the non-distributive lattices are `read off' from Hilbert spaces---complex vector spaces with an inner product defined.  This doesn't explain the relationship Hilbert spaces have to the quantum world, or why the quantum world can be formalized through Hilbert spaces.  A logic based on such a formalism does not get any closer to a physical \emph{meaning}, and in fact may obscure certain physical notions rather than clarify them.  }  It is not the main purpose of this present paper to criticize what can be called the traditional attempts in this area, however a few comments can be mentioned.

First, the original attempt from Birkhoff and von Neumann \cite{Birkhoffetal1936} (which seems to be much of the basis for subsequent literature, e.g. in Putnam \cite[\S 10]{Putnam1979}) appears to be a discussion over \emph{static} properties, such as posterior outcomes of measurements.  For quantum programs, it makes little sense to base our logical calculus on the posterior (and static) aspects of event structures.  Rather, we will be using coherent unitary transformations describing temporal dynamics in our quantum algorithms---and thus they should appear as the primary focus in the proof system.  A few recent attempts have shown that incorporating a time index or dynamic modalities can begin to reconcile this issue.  While I find it relieving that these recent attempts have breathed fresh life into the subject, I have not been particularly convinced.

Briefly, I can mention what aspects of quantum computation these approaches have not captured but which \emph{should} be captured in a proof system for quantum programs.  Importantly, as Baltag and Smets \cite{Smetsetal2006} readily admit,\footnote{One can glean the motivation for the present approach from the admission: ``Finally, we should mention here some of the \emph{limitations of our approach,} which arise from our \emph{purely qualitative, logic-based} view of quantum information.  The quantitative aspects are thus neglected [\dots] by abstracting away from complex numbers, `phases' and probabilities.  [\dots] our programs will be `\emph{phase-free}.'  This is a serious limitation, as phase aspects are important in quantum computation; there are ways to re-introduce (relative) phases in our approach, but this gives rise to a much more complicated logic, so we will leave this development for future work.''  \cite[p. 495]{Smetsetal2006}~  The authors then rightly admit that a more complicated logic would result from a treatment of these factors, and that such a logic would require a tensor operator.  Presumably what is meant by this latter comment is that there must be a connective corresponding to the tensor product, which we will incorporate here into a proof system.} treatments of phase relations and measurement probabilities are lacking of representation in their system.  This is critical, since many quantum protocols involve \emph{interference}---which is perhaps most apparent in ``sandwiched'' applications of Hadamard gates $\mathbf{H}$ (a particularly useful unitary transformation).  

%The quantum lambda calculus approach developed in \cite{Tonder2004} seems like a very fruitful contribution to quantum computation.  It is unclear to me at this time whether such lambda expressions could also be applied to the sequent representation suggested here---although this would be ideal and should be explored (i.e. are parts of the formulation here equivalent to the more advanced exposition in \cite{Tonder2004}).\footnote{For example, I will take unitary transformations such as the Hadamard gate to be of critical logical importance in the sequent formulation, they are the basic logical operations.  In the lambda calculus developed by van Tonder, these appear to be constants in the quantum lambda calculus.}  

This work should also be contrasted with the many attempts which utilize category theory to logically characterize quantum information processing.  These attempts are at a different level of analysis from the approach discussed here.  Category theory provides an alternative foundation for mathematics, and thus it presumes a high amount of mathematical structure and sophistication additional to the basic level of linear algebra and circuit diagrams.  While sympathetic to these approaches, it is arguably worthwhile to approach the present subject from an alternative route---namely starting with the formalisms and diagrams that you will learn in a quantum information course from a physics department.  

%Category theory is not the language in which quantum information processing is introduced, in this author's experience.  

Both routes are justified, but at the very least there is sufficient justification to fill in the gap between circuit diagrams and linear algebra on the one hand, and category theory on the other.  The level of analysis here may contribute to filling in this gap, by explicitly linking physical concepts and familiar formal tools to a preliminary sequent system.  At the very least, for mathematically inclined philosophers of science and physicists, this article can help motivate the deeper logical considerations which may be found in the category theory approaches.  At most, this article makes more explicit what features sequent style logics for quantum programs should represent (e.g. phase).  Furthermore, if certain approaches do not capture these features, then this present analysis helps point out the importance for proponents to explain \emph{why} they do not capture these features.

As a preliminary intuition, a logic for quantum programs should abstract away from some operations but still preserve the intuitive nature of quantum algorithms and the resources involved.  In particular it should represent (i) the \emph{phases} of a superposition;  and (ii) it should represent measurements, amplitudes, and measurement probabilities.  My focus is on the conceptual representation of quantum computational procedures in a sequent-like formulation.  Intuitively, the deductions in proofs will be seen as roughly  equivalent expressions to the circuit model depictions of quantum protocols.

\subsection{Motivation for a Logic of Quantum Programs}

Using the terms algorithm and program as loosely equivalent, a quantum algorithm is an explicit step by step procedure using the formalism of quantum mechanics\footnote{Dirac notation and the Hilbert space representation is presumed, rather than the somewhat cumbersome wave mechanical formulation.} to achieve some computational task.  Some famous examples of quantum algorithms are the Deutsch-Josza and Shor algorithms---the purpose of the latter is to factor products of two large prime numbers efficiently through period finding, potentially enabling us to break for example the widely-used RSA cryptosystem.  The possible implementation of these type of powerful tasks on some physical architecture creates a demand for deeper philosophical analysis and practical development of quantum programs.

Much of the philosophical literature on quantum logic has unfortunately been on why its underlying partial Boolean algebra (PBA) is of the structure of non-distributive ortho-modular lattices, and drawing consequences from this non-classicality.  However, in the opinion of the present author, such attempts are misled based on the aforementioned lack of dynamics in the logics.  It is simply not very interesting to discuss quantum logics which are static.  Girard \cite[\S 1]{Girard2003} would perhaps even trace the problem to a more general one in the history of logic, in which it seems that the static denotational or semantic aspects have been much more developed than the dynamic, constructive, and proof-theoretic aspects of logic.  

Rather, we should focus on finding a \emph{constructive} approach to a quantum logic, enabling us to establish a correspondence between proofs in this logic and algorithms or programs.  Also, the so-called Curry-Howard correspondence extends to a correspondence between propositions and types in type theory.  See e.g. Nordstrom \emph{et al.} \cite[p. 4]{Nordstrometal1990}.  While it is not my intention to prove this correspondence for the rules to be discussed, or to give typed lambda calculus representations of the proposed sequent calculus, it is with this correspondence in mind that I approach the field of quantum logic through proof theory and the important physical concepts involved rather than the traditional analysis of lattices and Hilbert space structures.  

%\begin{quote}
%``A specification [of a problem to be solved in a program/what a program should do] is in type theory expressed as a set, the set of all correct programs satisfying the specification.  The programming process is the activity of finding and formulating a program which satisfies the specification.  In type theory, this means that the programmer constructs an element in the set which is expressed by the specification.'' \cite[p. 4]{Nordstrometal1990} 
%\end{quote}
%
%
%We can formulate this relation something like:
%
%\begin{equation}
%{proof\over proposition} :: {term \over type} :: {program \over specification}
%\end{equation}

Other approaches towards developing quantum programming languages, and related formal aspects of theoretical quantum computer science, are varied and fragmented yet developing quickly.  See for example the survey from Ying \emph{et al.} \cite{Yingetal2012}.  Some particularly interesting developments in this area with respect to logic are dynamic modal logics put forward by Smets and Baltag in order to capture important aspects of quantum information theory and quantum programming.  These are inspired by what they call the ``dynamic turn'' in logic.  \cite{Baltagetal2012}  I have also already mentioned the growing popularity of categorical quantum mechanics, which I have been growing more sympathetic to.

It should thus be mentioned that the project discussed here is not original, but it also doesn't seem to be directly in line with these other approaches.  For example Selesnick's project in \cite{Selesnick2003} also explores the proof-theoretic and type-theoretic foundations for quantum computing, although the conceptual and semantic discussion is less present than what I hope to achieve here.  It is my hope that the familiar reader will admit that certain treatments found in this present work are more conceptually accurate to what a quantum algorithm is supposed to do than the aforementioned work.  For example, with regard to a sequent representation of measurement, but also it is not clear yet at this stage whether the proof rules discussed are capable to be incorporated into previous work in this area.  It may also be the case that similar conceptually-oriented rules have been proposed elsewhere, but as of the time of writing I have not encountered them.  

So, while I am looking to develop the proof-theoretic side of quantum programs, I am motivated by intuitive semantics and formalisms from quantum computation and the circuit model more than I am by formal quantum logics (which in my experience are \emph{not} used in the typical exposition of quantum computation and quantum algorithms).  It should also be emphasized that this work does not represent a final solution for a logic of quantum programming, but hopefully helps fill the gap between what is taught in an introductory quantum information course and the logical approaches mentioned above.

\subsection{What are we Talking About?}

Before entering into a discussion, where formal sequent calculus will primarily be used, we need to first establish what kind of notation to use---and determine what exactly the notation should be representing.  It seems that the former is in some ways more difficult than the latter.  It would be very nice to be able to write quantum algorithms\footnote{Or the quantum aspects of a procedure, if there are also classical tasks involved in an algorithm.} down without calculus, linear algebra, or vector notation (even though Dirac notation is quite nice).  However even then we would still like to have representations of phase and measurement probabilities, since these may be important aspects of the procedure---either for practical implementation or for theoretical understanding.  It seems that some mix of sequent calculus, Dirac notation, and linear algebra may end up being the most clear---and this is the route pursued here.  

While I cannot possibly hope to resolve the entire situation in this present paper, the motivation is clear:  if we are to have quantum algorithms, and implement them and discuss them in philosophy and theoretical computer science, then a `higher-order language' involving symbolic logic is preferable to working with differential equations and vector spaces.  In a logic for quantum programs, then, we should try to find a balance between compact formal logical representations of states, unitary operations, and measurements---without compromising the mathematical expressiveness that is necessary to understand \emph{how} quantum algorithms work.  For example, it seems that in our logic for quantum programs we should require that the Born Rule and its associated probabilities are represented---but I have not seen such distributions in the sequent-style representations put forward elsewhere (e.g. in Selesnick \cite{Selesnick2003}).  

As far as the language which the specification/propositions/types will range over, I consider it generally to consist of physical statements about the quantum world.  Typically these would be formulated in wave mechanics or equivalently in matrix mechanics (however the wave mechanics is more `ontic' with respect to the imagined token operation performing some type of task).  These statements should be about what happens during unitary transformations and the dynamics of the physical world as well as statements of  measurements.  Importantly, then, the statements shall not be limited to what Birkhoff and von Neumann originally called \emph{experimental} propositions (which would appear to be a subset of the possible \emph{physical} propositions).  

On a superficial level, the background language here can just be considered linear algebra.  Logical operations will be abbreviations of, for example, the tensor products of vectors and matrices.  With this in mind, we begin in the next section with a sequent analysis of representing and operating on these statements.

\section{Reading Sequents as Quantum Computations}

In the field of theoretical quantum computation, the following expression represents the quantum analogue of a binary digit or \emph{bit}, the \emph{qubit}:
\begin{equation}\label{qubit}
\ket{\psi} = \alpha\ket{0} + \beta \ket{1}.
\end{equation}

Here, $\ket{\psi}$ (read as ``ket psi'' in what is known as Dirac notation) is a superposition, i.e. a linear combination represented by the addition sign, of the two basis states $\ket{0}$ and $\ket{1}$.  Kets are taken to be column vectors whose entries are  complex numbers, whereas `bras' $\bra{\cdot}$ are row vectors whose entries are the complex conjugates of the corresponding ket.  For example, our basis states are the following column vectors:
\begin{equation}
{
\ket{0} = 
\begin{bmatrix}
1 \\
0 \\
\end{bmatrix}
},
{
\hspace{2cm}
}
{
\ket{1} = 
\begin{bmatrix}
0 \\
1 \\
\end{bmatrix}.
}
\end{equation}

The coefficients $\alpha, \beta$ in \eqref{qubit}  are complex numbers called \emph{amplitudes}, and this title is best understood through the wave mechanical formalism of quantum mechanics.  We require of these amplitudes that $|\alpha|^2 + |\beta|^2 = 1$, where $|\cdot|$ denotes the modulus of the complex number.  Such a calculation is known as the Born Rule, to be discussed further in due time.

From combinations of these building blocks, including unitary linear operators on state vectors and vector spaces containing them, linear algebra forms a formal language which is capable of representing procedures useful in theoretical computer science.  A sequent calculus based on these formal aspects will informally be read with this background in mind.  Furthermore, intuitive considerations from the field of quantum computation will help determine the types of rules to be used in the system, as well as how to properly read and interpret the applications of the rules.  

\subsection{Quantum Circuits}

The primary goal for the sequent system to be discussed in the following sections is to represent the quantum circuit model of quantum computation.  In this model, protocols are represented diagrammatically in  `circuits'.  As an example, consider the following `circuit':\footnote{The double lines after measurement are supposed to represent that we have a classical state, not a quantum one.}
\begin{equation}\label{entangledcircuit}
{
\includegraphics[width=8cm, height=2.5cm]{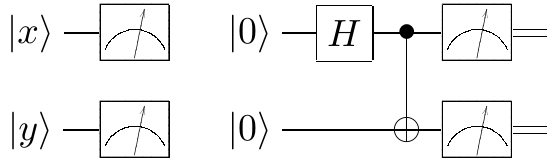}
}
\end{equation}

Where $\ket{x}$ and $\ket{y}$ are unknown states of single qubits, put through a measurement (represented by the box with the arrow).  Say both of these detectors display 0.  Then, as discussed in more detail shortly, we can assign the state $\ket{0}$ to both of these qubits.  This is called \emph{state preparation}, and will be used to define axioms in our sequent system---allowing us to cut the left part of this diagram off (which one could also do in the circuit diagram).  Then, in the subsequent part of the diagram, two transformations are performed on these states---the first is one of the most common gates in quantum computation called a Hadamard gate $\mathbf{H}$, represented by the $2\times2$ matrix
\begin{equation}
\mathbf{H} = {1\over \sqrt{2}} 
\begin{bmatrix}
1 & 1 \\
1 & -1 \\
\end{bmatrix}.
\end{equation}

Subsequently, the operation represented by the vertical link represents a two-qubit gate $\mathbf{CNOT}$, or controlled-not.  Depending on the state of the control qubit (where the black dot is), this $4\times4$ matrix will XOR (addition modulo 2) the second qubit.  
\begin{equation}
\mathbf{CNOT} =
\begin{bmatrix}
1 & 0 & 0 & 0 \\
0 & 1 & 0 & 0 \\
0 & 0 & 0 & 1 \\
0 & 0 & 1 & 0 \\
\end{bmatrix}.
\end{equation}

This is remembered since the normal notation for addition modulo 2 in this context is represented by $\oplus$, but this should not be confused with the later use of the same symbol to denote exclusive disjunction.  Whereas the first set of measurement gates was for state preparation, the second set on the right side of the circuit will be upon the two-qubit entangled (non-factorizable) state $\frac{1}{\sqrt{2}} \ket{00} + \frac{1}{\sqrt{2}} \ket{11}$, which the reader can check with the above statement of Born's rule will  result in measuring either $\ket{00}$ or $\ket{11}$ with the equal probability of one-half.

With this preliminary crash course in the formalism used in quantum computation and the circuit model of representing quantum protocols, we can now begin to outline a sequent calculus built to represent similar functions in quantum computer science.  We will see how this same entanglement-generating circuit is represented in a sequent calculus for single quantum circuits.

\section{Building a Sequent Calculus for Single Quantum Circuits}

\subsection{Axiom and State Preparation}

Given an unknown qubit state we can subject it to measurement in an appropriate basis, and by noting whether or not our measurement device shows a `0' or `1', we can assign the post-measurement state to either the basis state $\ket{0}$ or $\ket{1}$ (see Mermin \cite[\S 1.10]{Mermin}).  We can apply the $\mathbf{X}$ gate to the qubits in state $\ket{1}$ and thus prepare all of our states uniformly into $\ket{0}$ states.  This initialization procedure can be seen as giving us the `premises' of our sequent proofs, i.e. the leaf nodes will be initialized into the $\ket{0}$ state.  The following pure state can be considered effectively as an \emph{axiom} in our system:
\begin{equation*}
\AXC{$\ket{0} \seq $}
\RLab{$Ax$}
\DP
\end{equation*}

Although, given the above statements, such a state may be the conclusion of a measurement in a sub-derivation higher up in our proof tree.  This will be returned to later.  It will be useful to note that we will be able to initialize $n$ of these basis states by $n-1$ applications of the conjunction rule $\otimes$, discussed in the next section.  The reason for keeping the succedent blank will also be explained in due course, but for those who cannot wait, the way we will read this axiom is actually 
\begin{equation*}
\AXC{$\qax P(\ket{0})$}
\DP
\end{equation*}

\noindent where $P$ represents a probability distribution calculated through the Born Rule ($BR$).  For this pure state, $P(\ket{0}) = 1$.  We will want to `normalize' the introduction of this probability-calculation (represented by the predicate $P$), by putting it off until the penultimate step of the proof (i.e. the step which is the premise of a concluding measurement rule application).  From now on, my choice of notation of keeping the succedent empty will reflect this `normal' form.  One can see, though, that this isn't too much of a stretch from the usual axiom $A\seq A$, and one will see that it is quite natural given the measurement conditional reading of the sequent arrow, to be discussed in Section~\ref{sec:measurementsequentarrow}.  These initialized states are the primary inputs of our unitary transformations, which will be considered our quantum logic gates---or, in other words, the logical operations that introduce or eliminate other formal symbols.  

While the notation in the premise of the following rule has not yet been discussed I list it here for its relevance to the axiom above.  In general, after we make a measurement we read out some classical value.  If we were to measure again we would obtain the same result, leading us to ascribe the outcome of the measurement as the current state of the system:
\begin{equation*}
\AXC{$\Sigma \vdash_p \ket{x}_n$}
\LLab{$(Prep)$}
\UIC{$\ket{x}_n \seq $}
\DP
\end{equation*}

$\Sigma$ is used instead of $\Gamma$, as a reminder that it can be regarded generally as the superposition of $n$ qubits:\footnote{Superpositions will be denoted by the symbol `+'.  In the system considered here, it is introduced via use of a Hadamard gate, as a byproduct rather than a `true' connective with an introduction and elimination rule (it is `eliminated' when total destructive interference occurs, see Section~\ref{sec:examples} for more).  It will be important to note later, however, that it is precisely the `+' in a superposition which will give us the conjunctive-then-disjunctive sense on the meta level of reading our quantum sequent calculus.}
\begin{equation}\label{gensup}
\sum_{0\leq x \leq 2^n} \alpha_x \ket{x}_n,
\end{equation}

\noindent which would mean that the corresponding circuit being represented in sequent form was an $n$ qubit circuit, and that there were $n$ measurements leading subsequently to $n$ prepared qubits.  

\subsection{Terms in the Antecedent}

If $\Gamma$ (i.e. the entire antecedent) were non-empty with more than one $\Sigma$, how would we `read' the antecedent of the sequent?  Suppose we wrote the following:
\begin{equation*}
\AXC{$\ket{0}, \ket{0} \seq $}
\DP
\end{equation*}

How would we interpret the structural comma for a logic whose purpose is to describe or represent computational procedures in a quantum protocol---specifically a \emph{single} circuit?  In other words, what \emph{state} is the `computer' in?  For a constructivist account of the antecedent above, how did we arrive at $\ket{0}, \ket{0}$?  The commas will not play a role in a single circuit calculus, but return to denote the separation between multiple circuits in a multiple circuit calculus (which will not be fully presented in this present short article).  Diagrammatically, one can easily see the differences between single and multiple circuits:
\begin{equation}
\includegraphics[width=4cm, height=5cm]{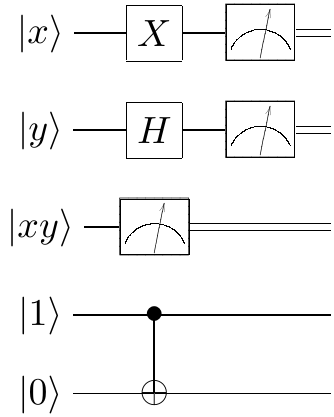}
\end{equation}

Here, $\ket{x}$ and $\ket{y}$ are single qubits each in their own separate circuits---the gates $\mathbf{X}$ and $\mathbf{H}$ are single qubit gates.  For now, though, we are concerned with a single circuit sequent calculus, which can be thought of as a way to incorporate multiple circuits into a larger single circuit.  Let us first take a look at some typical left conjunction rules, since both states must be in the same circuit.

\subsubsection{Multiplicative vs. Additive AND}

The following is the intuitionistic rule (in $\mathbf{G1ip}$, the same rule as $\mathbf{G1cp}$ but restricted in the succedent) for introducing $\land$ in the antecedent:
\begin{equation*}
\AXC{$\Gamma, A_i \seq C$}
\RLab{$(L\land)(i \in \{ 0, 1 \})$}
\UIC{$\Gamma, A_0 \land A_1 \seq C$}
\DP
\end{equation*} 

In a system like $\mathbf{G1ip}$ with left weakening (and left contraction), the above `additive' rule can be shown equivalent to its `multiplicative' counterpart:
\begin{equation*}
\AXC{$\Gamma, A_0 , A_1 \seq C$}
\RLab{$(L \land ')$}
\UIC{$\Gamma, A_0 \land A_1 \seq C$}
\DP
\end{equation*}

This multiplicative conjunction is obviously the type of conjunction that can help us create a single circuit from $\ket{0}, \ket{0}$ simply because both circuits must be present in order to combine them.  However, it will be noted in further detail later on that we should \emph{not} generally admit left weakening into a proof system for quantum programs because of interference---and so there is no such equivalence between additive and multiplicative conjunctions.  There is no additive conjunction.  In this case, it may be better not to use the $\land$ symbol to represent our multiplicative conjunction.  From Girard's linear logic, we could also use the conjunctive symbol `$\otimes$' for our conjunction:
\begin{equation*}
\AXC{$\Gamma, A_0 , A_1 \seq C$}
\RLab{$(L \otimes)$}
\UIC{$\Gamma, A_0 \otimes A_1 \seq C$}
\DP
\end{equation*}

This symbol will be preferred, since it will be mimicking the tensor product in linear algebra.  While no tensor products will actually be calculated in the logic, such operations must be carried out in the background or by hand.  There will be no contexts $\Gamma$ in the single circuit calculus (but they return for multiple circuits).  We now have a proposal for a multiplicative conjunction to include in our system---call it \logicname ~(Quantum sequent calculus for Measurements and Circuits).  The system will also be intuitionistic in the sense that there will only ever be one formula in the succedent---the sub-formulas of which will be candidate formulas for succedents of conclusions in a quantum proof.\footnote{In other words, measurement will pick out some component from the superposition.}  More on this later.

Unfortunately there are still some unsatisfactory issues, which can perhaps be alleviated by the following constructive account of the terms in the antecedent.  Take for example the notation mentioned earlier of the terms denoted by $\ket{0}, \ket{0}$.  Let the following `rule' $LComma$ construct such a term from two instances of our axiom---but notice that the `conversion' on the right shows the equivalence between $L\otimes$ above and $L\otimes '$ in the presence of the comma introduction rule:
\begin{center}
{
\AXC{$\qax $}
\AXC{$\qax  $}
\LLab{$(LComma)$}
\BIC{$\ket{0}, \ket{0} \seq $}
\LLab{$(L\otimes)$}
\UIC{$\ket{0}\otimes \ket{0} \seq $}
\DP
}
{
$\rightsquigarrow$
}
{
\AXC{$\qax  $}
\AXC{$\qax  $}
\LLab{$(L\otimes ')$}
\BIC{$\ket{0}\otimes \ket{0} \seq $}
\DP
}
\end{center}

The rule $L\otimes '$ is considered more primitive since it not only captures the constructive, multiplicative, and conjunctive aspect of our resources---it also avoids the $LComma$ rule application.  Since we will not have a $R\otimes$ rule, I will refer to $L\otimes '$ as $\otimes$ in \logicname.   Using this rule, we can initialize (or represent the initialization of) $n$ $\ket{0}$ states (written by convention $\ket{0}^{\otimes n}$) by the application of $n-1$ gates, or rule applications, of $\otimes$.\footnote{Convention in the quantum computational literature for tensor products (here mimicked by the same  symbol $\otimes$) of $n$ states is $\ket{0}^{\otimes n} = \ket{0}_1 \otimes \dots \otimes \ket{0}_n = \ket{0}_1 \ket{0}_2 \dots \ket{0}_n = \ket{0_1 \dots 0_n}$.}

%In the general case it takes the form
%
%\begin{center}
%\AXC{$\Sigma_n \seq$}
%\AXC{$\Sigma_m \seq$}
%\LLab{$(\otimes)$}
%\BIC{$\Sigma_n \otimes \Sigma_m \seq $}
%\DP
%\end{center}

\subsection{Measurement as the Sequent Arrow}\label{sec:measurementsequentarrow}

Hughes \cite[p. 303]{Hughes1989} discusses the notion of a ``measurement conditional'', formulated in terms of the probability of a quantum observable $A$.  This is to be read ``If an ideal measurement $M$ of $A$ is made, then the result will lie within some range $\Delta$.''  
\begin{equation}
MA \to (A, \Delta).
\end{equation}

However, for the purposes here it will be helpful to instead formulate measurement as a proof rule rather than a predicate associated with a connective.  In other words, there will not be an arrow connective in \logicname.  Rather, the sequent arrow will be read in the way that the above measurement conditional arrow is read.  While the counterfactual reading of the measurement conditional will be retained in the sequent calculus used here, a proof ending in an actual measurement cannot continue in this manner since a measurement \emph{has been performed}.  Thus, the reading of the sequent arrow $\seq$ will switch from a counterfactual one to a statement of what has been proved (measured).  The principle of deferred measurement in the circuit model of quantum computation states that we can always put off our measurements to the end of whatever computational procedure we wish to perform.  That is, in cases where we do not use classical information extracted from the measurement (as in the teleportation protocol), we can defer the measurements to a later stage while preserving the essential elements of the algorithm.  

The result of a quantum program will depend on measurements of the state, in which we `call' the result of a sequence of unitary transformations (the program).  It is at this point that the `conclusion' of the proof, the outcome of the program, is reported.  Very generally, and again this is part of a conceptual approach being developed here, we can regard the sequent arrow `$\seq$' as indicating a relation between a potential quantum measurement on states and the probability distribution that characterizes the outcome.  Antecedent terms discussed in the previous section are taken to be coherent states of computations in a quantum computer, whereas the terms in the succedent will represent possible outcomes after evaluating the result of the computations.

When we view the sequent arrow in this manner, we `bake' into the syntax a dynamic aspect.  This can be interpreted as representing in one sense \emph{time}, and in another sense the non-unitary character of a measurement process (either epistemically or literally depending on the interpretation).  The temporal aspect is that the potential outcomes of measurements, catalogued in the succedent, are presumed to occur at some time after a measurement.  In other words, the succedent would be later in time than the antecedent if the proof ended and a measurement was made.  Additionally, we may want to characterize the conclusion of each unitary logic gate as being at a later time than the line before---i.e. each line is at a later time than the previous line.\footnote{This temporal sense of a single circuit must be carefully applied in a multiple circuit calculus.}  

%A further dynamic aspect is the essentialy dynamic calculation of measurement probabilities---which changes every time we apply unitary transformations to part or all of the coherent state in the antecedent.  

\subsection{Updating the Born Rule and Disjunction in the Succedent}

To make sense of quantum measurements, Max Born famously proposed we take the complex-valued \emph{amplitudes} $\alpha, \beta$ of wave components in a state and calculate the modulus squared of each such that, to put it simply, $|\alpha|^2 + |\beta|^2 = 1$.  In the measurement of, for example, the state $\alpha\ket{0} + \beta\ket{1}$, we will only observe \emph{either} the state represented by $\ket{0}$ \emph{or} the state represented by $\ket{1}$---but with corresponding probabilities determined by the Born Rule process.  Since will only observe one of the possible components in the superposition, I take this to be an \emph{exclusive} sense of the disjunction.  

%\begin{quote}
%``Measurement gates therefore play two roles in a quantum computation. They get the Qbits ready for the subsequent action of the computer, and they extract from the Qbits a digital output after the computer has acted. The initial action of the measurement gates is called state preparation, since the Qbits emerging from the process can be characterized by a definite state. The association of unitary operators with the gates that subsequently act on the Qbits permits one to update that initial state assignment into the corresponding unitary transformation of the initial state, thereby making it possible to calculate, using the Born rule, the probabilities of the outcomes of the final measurement gates.''\cite[p. 31]{Mermin}
%\end{quote}

In the project here, two types of calculations in particular are relegated to background sub-routines (i.e. the formalism here abstracts away from them).  These are the basic linear algebraic manipulations such as evaluating tensor products of matrices, but also the calculation through Born's rule of measurement probabilities.  The introduction of a predicate $P$ into the succedent represents this on a conceptual level, but of course the distribution must be calculated for specific examples (i.e. specific $\Sigma$):
\begin{equation*}
\AXC{$\Sigma \seq$}
\LLab{$(BR)$}
\UIC{$\Sigma \seq P(\Sigma)$}
\DP
\end{equation*}

In general, the probability distribution introduced by $BR$ would need to be updated after every application of a unitary gate.  However, we aren't necessarily concerned with such calculations until the time when we \emph{actually} want to make a measurement.  I propose the following sequent rule for measurement:
\begin{equation*}
\AXC{$\Sigma \seq P(\Sigma)$}
\UIC{$\Sigma \seq \ket{x}_n$}
\DP
\end{equation*}

\noindent where again $\Sigma$ can be regarded as containing the general superposition in \eqref{gensup}.  We should note that in this rule the sequent arrow has completely changed from a potential indicator of possibilities to an actual statement of what the state $\Sigma$ has ended up being \emph{actually} measured as.  An alternative formulation in which the conclusion is represented instead by the turnstile is warranted:
\begin{equation*}
\AXC{$\Sigma \seq P(\Sigma)$}
\LLab{$(M)$}
\UIC{$\Sigma \vdash_p \ket{x}_n$}
\DP
\end{equation*}

Then we can still read the penultimate sequent arrow in the counterfactual sense found in Hughes' measurement conditional, while reading the ~$\vdash$~ as `proves' or `measures', with probability $p$ as a subscript.  If part of the definition of an intuitionistic system is that there is only one term in the succedent, then this system must be intuitionistic in this sense.   There is not an $M$ application which results in more than one term in the succedent.  

As a final note in this section, weakening on the \emph{right} is disallowed on the intuitive basis that either (i) before measurement we cannot extend the Born rule (i.e. our probability distribution) over more states in the succedent than are present in the antecedent (i.e. in the coherent quantum state); or (ii) after measurement our derivation (or sub-derivation) is finished, and the only rule allowing us to continue from the turnstile form is $Prep$ which moves the measured state into the antecedent.

\section{Interference and Non-monotonicity}

The addition of a new state during our computation will \emph{interfere} with the other states in a superposition.  This can be viewed as a desirable representational feature of the calculus.  One of the primary features of a quantum computer, acknowledged in the literature as a potential mechanism or resource enabling so-called quantum speed up, is quantum interference (see e.g. Fortnow \cite{Fortnow2003}).  Interference allows certain `computational paths' to destructively interfere (i.e. two equal, oppositely signed amplitudes of the same state will make a term disappear).  

An important feature which this proof system should then exhibit is that adding `premises' (or states in the antecedent) does not automatically guarantee that the same terms in the succedent are available to be measured or proved.  This is an important logical feature, called non-monotonicity, and for many modern logicians such a property is  seen as a beneficial representational aspect of a logical system.  For example, it can reflect that upon adding new information to our knowledge or set of beliefs, the conclusions we might draw from reasoning with this set might change.  Monotony would imply that adding elements to a premise set would still entail the same conclusions.  In sequent notation, we can see that left weakening closely represents the monotonic notion (here presented with a general turnstile, not necessarily the post-measurement turnstile used previously):
\begin{equation*}
\AXC{$\Gamma \vdash \Delta$}
\LLab{$(LWeak/Mon)$}
\UIC{$\Gamma, \Gamma' \vdash \Delta $}
\DP
\end{equation*}

Thus, we can intuitively explain why our system will not in general admit of left weakening---interference implies that the sequent consequence arrow (interpreted in terms of measurement as discussed above) is \emph{non-monotonic}.  That a quantum proof system should not generally admit weakening because of interference has also been noted by Selesnick \cite[p. 406]{Selesnick2003}, and perhaps elsewhere.  For a single-circuit sequent calculus, we have already discussed that the comma here must be interpreted as (and replaced by) the conjunctive connective $\otimes$.\footnote{Following this note, one can see that unlike in the classical case the set $\{a, b\}$ is not equivalent as a premise to the conjunction $a\otimes b$.  Makinson \cite[p. 6]{Makinson2005} calls it `conjunctive' monotony, when the set does behave the same as the conjunction.}  There are, however, several formulations of monotonicity.  One could call the following \emph{general} monotonicity:
\begin{equation*}
\AXC{$\Gamma \vdash \Delta$}
\LLab{$(GMon)$}
\UIC{$\Gamma, \Gamma' \vdash \Delta, \Delta' $}
\DP
\end{equation*}

This rule could be admissible (i.e. shown that it does not lead to new derivations) in a system with both left and right weakening.  For the present purposes, this is obviously not allowed in the quantum sequent system being considered.  Weakening is not allowed on either side.  General monotonicity returns in a sequent calculus for multiple circuits, under the assumption of non-interfering separate circuits.  However, the more fundamental discussion of sequent representations of single circuits here is needed first to understand the underlying properties, such as the relationship of interference to non-monotonicity.

\section{Examples and Results}\label{sec:examples}

In addition to the proof-theoretic rules that have so far been discussed, we can add a universal set of unitary transformations.  That is, there is a group of left rules $U$, where $U$ is ideally some universal set of unitary operators.  These rules are more difficult to represent in a general form, and in multi-partite operations their applications to specific qubits requires care.  The form of the Hadamard gate for example, for $x$ in the basis states $\{0, 1\}$, is:
\begin{equation*}
\AXC{$\ket{x} \seq $}
\LLab{$(\mathbf{H})$}
\UIC{$\frac{(-1)^x}{\sqrt{2}} \ket{x} + \frac{1}{\sqrt{2}}\ket{1 - x} \seq $} 
\DP
\end{equation*}

More work should be done on representing other unitary gates in the sequent formalism.  Applying unitary matrices to \emph{states} in quantum computation are the type of operations we want to represent in a quantum proof system.  In effect, the sequent representation proposed here is strikingly similar to quantum circuit representations by construction.  For example, to generate or initiate an entangled state we can construct the following sub-proof (or sub-program):
\begin{equation}
\AXC{$\qax$}
\LLab{$(\mathbf{H})$}
\UIC{$\hadz \seq$}
\AXC{$\qax$}
\LLab{$(\otimes)$}
\BIC{$(\hadz)\otimes \qax$}
\LLab{$(\mathbf{CNOT})$}
\UIC{$\frac{1}{\sqrt{2}} \ket{00} + \frac{1}{\sqrt{2}} \ket{11} \seq$}
\DP
\end{equation}
%\end{center}

%This is an unfinished program since we have not calculated measurement probabilities in the succedent, and have not applied a measurement rule $M$ yet.  There are many different ways to construct this particular entangled state.  However, such a coherent state will be useful as an input for other computational steps.  In the background, the state written as the conclusion of $\otimes$ actually is an example of distributivity.  In other words, $(\ket{x} + \ket{y}) \otimes \ket{z} = \ket{x} \otimes \ket{z} + \ket{y} \otimes \ket{z}$.  (The $\mathbf{CNOT}$ application above flipped the $\ket{0}$ above in the second $\ket{z}$ position to a $\ket{1}$.)  This illustrates that on the surface of this proof system, we regard $\otimes$ as a more or less normal connective satisfying distributivity (but not commutativity), since in the background calculations performed it is distributive (linear) over the superposition `$+$' of vector components.  

Continuing the above sub-derivation we can conclude the proof with an application of $BR$ and a measurement.  Remembering that $\ket{00} = \ket{0}\ket{0}$, we can use letters as subscripts to denote Alice's and Bob's registers (useful for example in discussing cryptographic protocols):
%\begin{center}
\begin{equation}
\AXC{$\frac{1}{\sqrt{2}} \ket{00}_{AB} + \frac{1}{\sqrt{2}} \ket{11}_{AB} \seq$}
\LLab{$(BR)$}
\UIC{$\frac{1}{\sqrt{2}} \ket{00}_{AB} + \frac{1}{\sqrt{2}} \ket{11}_{AB} \seq \frac{1}{2} \ket{00}_{AB} + \frac{1}{2} \ket{11}_{AB}$}
\LLab{$(M)$}
\UIC{$\frac{1}{\sqrt{2}} \ket{00}_{AB} + \frac{1}{\sqrt{2}} \ket{11}_{AB} \vdash_{\frac{1}{2}} \ket{00}_{AB}$}
\DP
\end{equation}
%\end{center}

The other proof available with equal probability would end in $\ket{11}_{AB}$ in the succedent of the conclusion of the measurement application.  In either case, there is perfect correlation between Alice's and Bob's registers for this entangled state.

\subsection{Interference as a Resource in Quantum Computation}

%In the constructive approach to quantum algorithms, here represented by step-by-step proof rules and background sub-routines evaluating linear algebra, we find an interesting result concerning the potential physical mechanism or resource used in the theoretical quantum computer.  Not only are entangled states constructed (i.e. not primitive), we get an insight into perhaps the primary resource involved in a quantum protocol.  

\emph{Interference}, while acknowledged in the literature, seems to be seen as less important of a resource than entanglement.  Both entanglement and interference seem to be important, since we see that entangled states are constructed (i.e. not primitive).  Another example will help to show a case in which interference is prominent.  Two applications of the Hadamard gate $\mathbf{H}$ creates two oppositely phased components of the same state in the superposition.  
%\begin{center}
\begin{equation} \label{HH}
\AXC{$\qax$}
\LLab{($\mathbf{H}$)}
\UIC{$\hadz \seq$}
\LLab{$(\mathbf{H})$}
\UIC{$\frac{1}{2}\ket{0} + \frac{1}{2}\ket{1} + \frac{1}{2}\ket{0} - \frac{1}{2}\ket{1} \seq $}
\DP
\end{equation}
%\end{center}

Again, background calculations have been performed by a sub-routine (in this case, by hand).  One can see, though, that the two $\ket{1}$ components when added together cancel out, leaving us with an amplitude (and subsequent probability) of 1 for $\ket{0}$.  In other words, on the last line the register is simply in state $\ket{0}$.  This ``Hadamard-sandwiching'' technique is used in a variety of quantum computational protocols, including some of the most powerful protocols such as Grover's and Shor's algorithms, as well as for ancillary qubits used in error correction (see e.g. Mermin \cite{Mermin}).  I have written out the interfering state explicitly for the reader to see interference in action---to see why phase is an important aspect of logical representation in quantum information science.

\bibliographystyle{eptcs}
\bibliography{SCRQCbib}
\end{document}